\documentclass{ws-procs9x6}

\begin{document}

\title{Results of Dark Matter Searches with the MACRO experiment}

\author{I. De Mitri}

\address{Dipartimento di Fisica,  Universit\`a di Lecce  and INFN \\
Via per Arnesano , I-73100 Lecce, Italy\\ 
E-mail: ivan.demitri@le.infn.it \\
- \\
On behalf of the MACRO Collaboration 
\footnote{See [\protect\cite{tecpap}] for the complete author list. } }

\maketitle

\abstracts{ The results of dark matter searches with the MACRO experiment are reported.
In particular indirect searches for WIMP's and direct searches 
for supermassive GUT magnetic monopoles are reported together with 
massive neutrino studies through the measurement of the oscillation 
induced anomalies in the atmospheric neutrino flux. }

\section{Introduction}
\label{sec:intro}
MACRO was a large multipurpose underground detector located in the Hall B 
of the Laboratori Nazionali del Gran Sasso.
The detector, which took data until December 2000, was arranged in a  
modular structure of six ``supermodules" with
global dimensions of $76.5 \times 12 \times 9.3\,$m$^3$ and a 
total acceptance  of $\sim 10,000$ m$^2$sr to an isotropic flux of
particles.
It was made by several layers of liquid scintillation counters,
limited streamer tubes and nuclear track detectors (CR39 and Lexan) which provided
excellent timing ($\sigma_t \sim 500 ps$), tracking ($\sigma_{\theta} \sim 0.5^\circ$),
and energy loss measurements.
For a complete technical description of the detector see [\cite{tecpap}].

\section{GUT Monopole searches}
\label{sec:mono}
In the framework of Grand Unified Theories (GUT), magnetic 
monopoles would have been produced in the very early Universe
as supermassive ($M \sim 10^{17} \,$GeV/c$^2$) topological defects [\cite{Preskill}]. 
One of the primary aims of the MACRO experiment was the search for such particles
with a sensitivity well below the Parker bound 
(i.e. $10^{-15}$ cm$^{-2}$ s$^{-1}$ sr$^{-1}$ [\cite{Turner82}])
in a very wide velocity range $4 \cdot 10^{-5}<\beta<1$ ($\beta=v/c$).

Different hardware systems were designed and operated to give
optimum sensitivity  in different $\beta$ ranges.
Moreover several analysis strategies were adopted depending on the 
used subdetector(s).
This approach ensured redundancy of information, efficient background rejection  
and independent signatures for possible monopole candidates.

\begin{figure}[ht]
\vskip -3.5cm
\centerline{\epsfxsize=3.9in\epsfbox{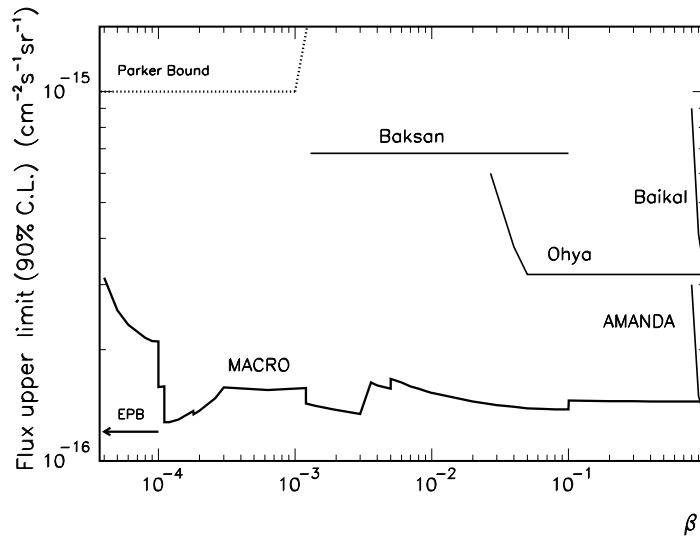}}   
\caption{\label{fig:limiti}
          The global 90\% C.L. upper limits to an isotropic flux 
          of bare magnetic monopoles obtained by MACRO and by other experiments
          [\protect\cite{Alexeyev90,Orito91,Amanda}] in direct searches (see text).}
\end{figure}

The analyses were based on the various subdetectors in a
stand-alone and/or in a combined way. Final results are reported in 
[\cite{monofin}] where all the details are given.
Since no candidates were detected, the results are given in as 
$90\%$ C.L. upper limits to the monopole flux (see Fig.\ref{fig:limiti}).
In order to compare the MACRO results to those of other experiments or 
to theoretical models, we present upper limits for an isotropic flux of bare MMs with
magnetic charge equal to one Dirac charge $g=g_D=e/2\alpha$,
(where $e$ is the electron charge and $\alpha$ the fine structure
constant) and nucleon decay catalysis cross section smaller than 1$\,$mb
[\cite{Preskill}].
A dedicate analysis for the search for magnetic monopoles accompanied 
by one or more nucleon decays along their path was also performed, resulting in very 
strong upper limits [\cite{catalysis}].

\section{Atmospheric neutrino oscillations}
\label{sec:atmo}
\begin{figure}
\vspace{-0.5cm}
\centerline{\epsfxsize=2.7in\epsfbox{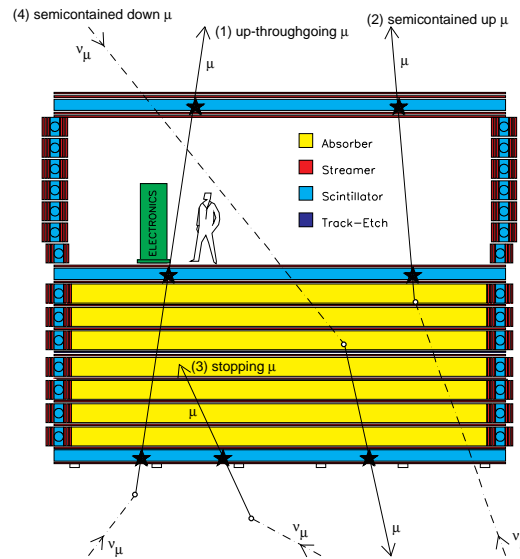}}   
\caption{\label{fig:topo}\small
  Topologies of events induced by neutrino interactions inside or around
  the detector (see text). The energy threshold for a muon to cross the detector 
  is $\simeq 1\,$GeV.}
\end{figure}

The study of the atmospheric neutrino oscillation was a primary goal
of the MACRO experiment. The streamer tubes and the scintillator counters 
allowed to distinguish neutrino events on the basis of time-of-flight 
($ToF$) measurement and/or topological criteria.
In particular four different types of neutrino events have been detected in MACRO.
(see Fig.~\ref{fig:topo}), corresponding to different topologies.
In the last two, i.e. {\it (3)} and {\it (4)}, the muons cross just one 
scintillator plane and then a $ToF$ measurement can not be done. 
Since it is impossible to distinguish between them, they are studied together as 
the sample {\it (3)+(4)}.

The parent neutrino energy spectra are of course different for different event types.
In particular the sample {\it (1)} is due to neutrinos with an average energy of
$\sim 50 \,$GeV, while other subsamples have lower energies ($< E_\mu > \simeq 4\,$GeV). 
Moreover the spectrum and the median energy of {\it (2)} and {\it (3)+(4)} are almost the
same.

Information on neutrino masses and oscillations have been obtained from the study of the 
angular distribution of the events of sample {\it (1)} and independent cross checks have 
been obtained from the other samples and from the measured event rates [\cite{upmu,nulowe}].

For the events in sample {\it (1)}, both the shape of the angular distribution and 
the absolute rate have been independently tested with the hypothesis of neutrino 
oscillation. 
The largest probability is obtained in the $\nu_\mu \rightarrow \nu_\tau$ oscillation 
scenario with maximal mixing and $\Delta m^2 \simeq 2.5 \cdot 10^{-3} \,$ eV$^2$,
while the $\nu_\mu \rightarrow \nu_{sterile}$ [\cite{nuster}] and the no-oscillation 
hypothesis are strongly disfavoured.
These conclusions are also confirmed by the analysis of the angular distribution and absolute
event rates of the low energy samples (i.e. {\it (2)} and {\it (3)+(4) }) [\cite{nuleproc}].

Moreover the different energy spectra of variuos event categories and the setting up of a 
method for the muon energy estimate [\cite{mcstec}] provided another independent tool for the 
study of atmospheric neutrino oscillations. 
In particular the behaviour of the oscillation probability as a function of 
$L/E$ (i.e. the ratio of the neutrino pathlength to its energy) has been obtained and shown to be 
compatible with the oscillation scenario suggested by the analyses of the angular distributions and
absolute event rates [\cite{nuleproc}].

\begin{figure}
\vspace{-2.5cm}
\centerline{\epsfxsize=3.0in\epsfbox{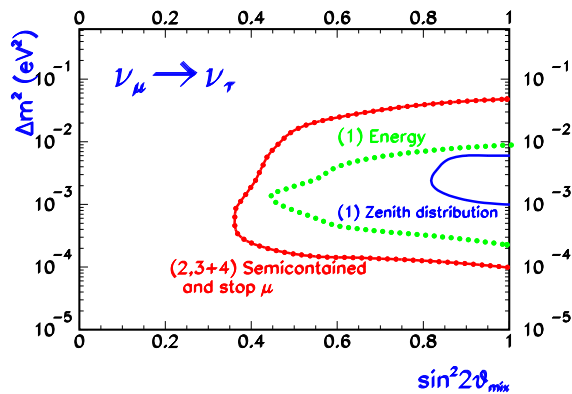}}   
\caption{  \label{fig:allow}\small
Allowed regions (Feldman-Cousins procedure [\protect\cite{feldcous}] with 90$\%$C.L.)
in the $\Delta m^2 - sin^2(2\theta)$ plane for $\nu_\mu \rightarrow \nu_\tau$ oscillation.
The continuos line is the result of the analysis of the angular distribution and absolute event 
rate for the sample {\it (1)}, while the dotted line refer to the analysis of the same data by using
the muon energy estimate. The last line is the same as the continuous one but refers to samples
{\it (2)} and {\it (3)+(4)} .}
\end{figure}

The results of all these analyses are shown in Fig.\ref{fig:allow} where the allowed regions in the
$\Delta m^2 - sin^2(2\theta)$ plane are shown for $\nu_\mu \rightarrow \nu_\tau$ oscillation. 
A global analysis to combine all this information is in progress.


\section{Indirect search for WIMPs}
\label{sec:wimps}
An indirect search for WIMPs has been performed by looking at 
neutrino-induced upward-going muons resulting from their annihilation in 
the Sun and/or the Earth [\cite{wimp,wimpnew}]. 

\begin{figure}
\begin{center}
\begin{tabular}{cc}
  \epsfxsize=2.2in\epsfbox{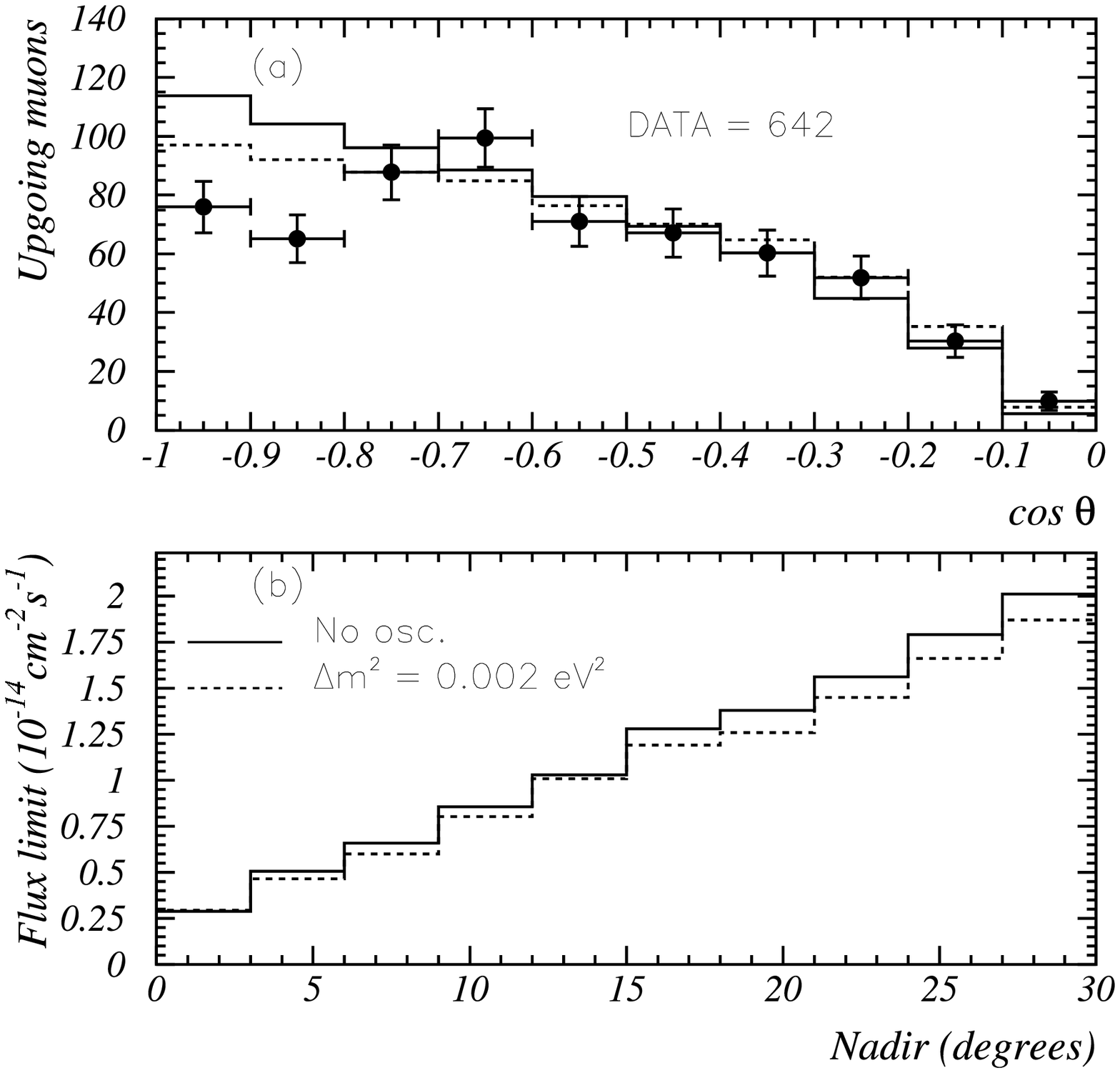}
  \epsfxsize=2.2in\epsfbox{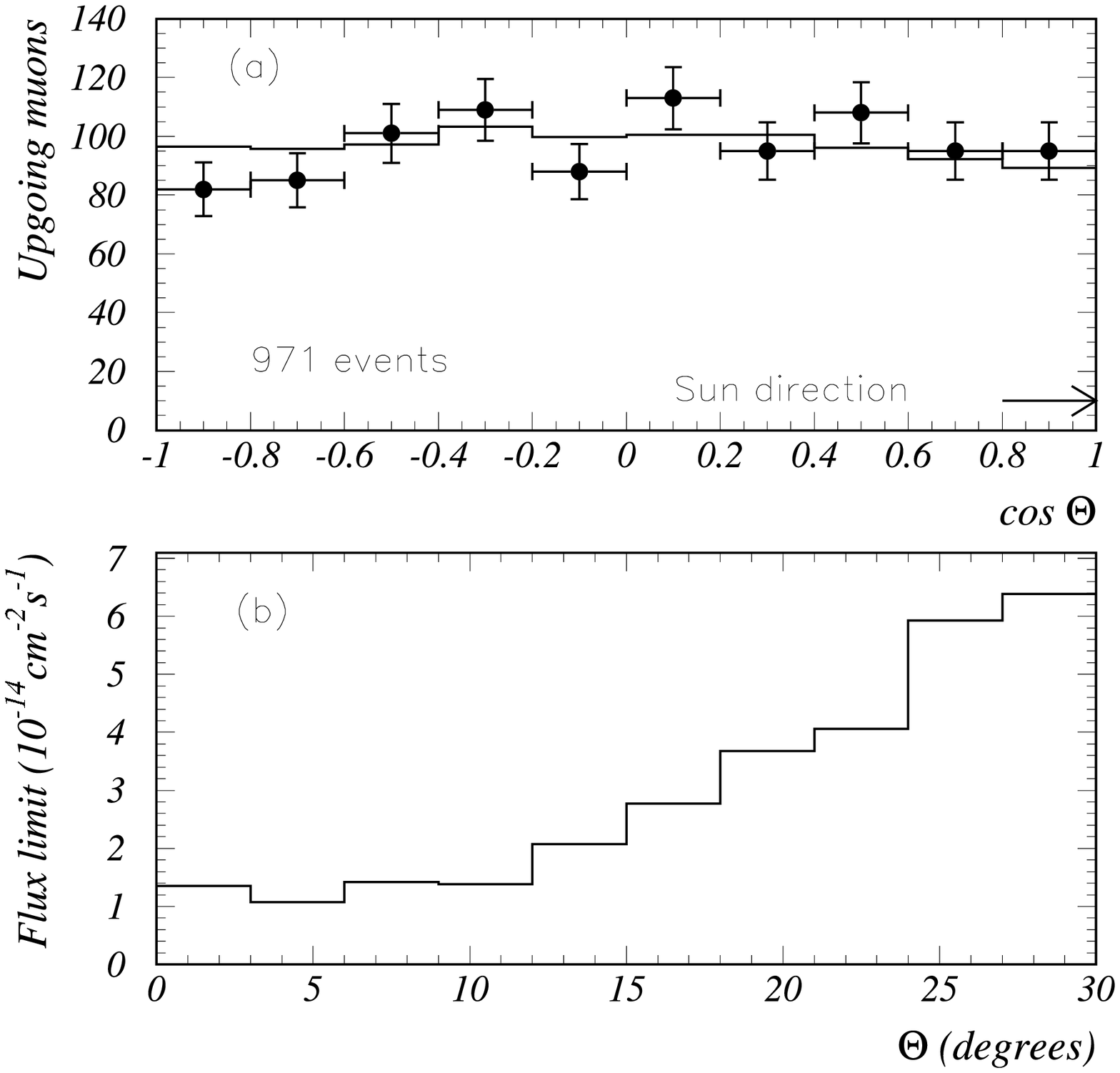}   
\end{tabular}
\end{center}
\caption{  \label{fig:wimp}\small
Results of indirect WIMP searches for the Earth (left) and the Sun (right). 
In the upper part the measured zenith angle distributions are compared with expectations 
for atmospheric $\nu$s in the case of no-oscillation (solid line) and $\nu_{\mu} \rightarrow \nu_{\tau}$
oscillation with $\Delta m^{2} = 0.0025$ eV$^2$ (dotted line).
In the lower part the $\mu$ flux limits (90$\%$ c.l.) vs. the half-width of the cone from 
the source direction are shown.}
\end{figure}

Muon flux limits have been obtained by comparing the measured values with the estimates given by detailed 
simulations taking into account the variuos background sources (i.e. the atmospheric neutrino flux, 
the backscattered soft pions produced by downward-going cosmic ray muons , etc.) and neutrino 
oscillation effects. 
This has been done for several search cones around the source (see Fig.\ref{fig:wimp}). 
A given search cone is ultimately related to the WIMP mass through a specific theoretical model.

The obtained limits can constrain the neutralino particle parameters suggested by SUSY models
[\cite{susywimp,summary}]. 

\section{Conclusions}
\label{sec:concl}
The MACRO results for dark matter searches have been reported.

Several searches for supermassive GUT monopoles have been performed 
by fully exploiting the redundancy and complementarity
offered by the detector together with the very large acceptance and the 
long time exposure. 
No candidates have been found thus putting an upper limit wich is well below 
both the Parker and closure bounds in a very wide velocity range. 
This excludes these particles as a main component of dark matter, by means 
of a direct experimental search. 

The atmospheric neutrino flux was measured by means of neutrino induced 
muons crossing the detector. Different event topologies were analyzed 
corresponding to different energy ranges. The results coherently favor a 
$\nu_\mu \rightarrow \nu_\tau$ oscillation scenario with respect to the case of 
$\nu_\mu \rightarrow \nu_{sterile}$ or to the no-oscillation hypothesis. 
Data are consistent with maximal mixing and $\Delta m^2 \simeq 2.5 \cdot 10^{-3} \,$ eV$^2$.

An indirect search for WIMP's has also been performed by looking at 
neutrino-induced upward-going muons resulting from their annihilation in 
the Sun and/or the Earth. Since no significant excess has been detected above 
the background, experimental upper limits have been set that can constrain 
the neutralino particle parameters suggested by SUSY models.


\begin{thebibliography}{}
\bibitem{tecpap}     M.Ambrosio et  al., {\it Nucl. Instr. \& Meth. in Phys. Res.} {\bf A486}, 663 (2002).
\bibitem{Preskill}   J.Preskill, {\it Ann. Rev. Sci.} {\bf 34}, 461 (1984) and references therein.
\bibitem{Turner82}   M.S.Turner, E.M.Parker, and T.J.Bogdan, {\it Phys. Rev.} {\bf D26}, 1926 (1982).
\bibitem{monofin}    M.Ambrosio et al., HEP-EX/0207020 (2002) , accepted by {\it European Journal of Physics C}.
\bibitem{catalysis}  M.Ambrosio et al., HEP-EX/0207024 (2002) , accepted by {\it European Journal of Physics C}.
\bibitem{Alexeyev90} E.N. Alexeyev et al.,(``Baksan''), $21^{st}$ ICRC, Adelaide,  vol. 10, 83 (1990).
\bibitem{Orito91}    S. Orito et al.,(``Ohya''), {\it Phys. Rev. Lett.} {\bf 66}, 1951 (1991).
\bibitem{Amanda}     G. Domogatsky for the Baikal Coll. (``Amanda'' and ``Baikal''),
                     XIX Int. Conf. on Neutrino Phys. and Astrophys., Sudbury, (2000).
%
%
\bibitem{upmu}       S.Ahlen et al , {\it Phys. Lett.} {\bf B357}, 481 (1995) \\
                     and M.Ambrosio et al., {\it Phys. Lett.} {\bf B434}, 451 (1998).
\bibitem{nulowe}     M.Ambrosio et al., {\it Phys. Lett.} {\bf B478}, 5 (2000).
\bibitem{nuster}     M.Ambrosio et al., {\it Phys. Lett.} {\bf B517}, 59 (2001). 
\bibitem{nuleproc}   M.Giorgini et al., HEP-EX/0210008 (2002).
\bibitem{mcstec}     M.Ambrosio et al., {\it Nucl. Instr. \& Meth. in Phys. Res.} {\bf A492}, 376 (2002).
\bibitem{feldcous}   G.J.Feldman and R.D.Cousins, {\it Phys. Rev} {\bf D57}, 3873 (1998).
%
%
\bibitem{back}       M.Ambrosio et al., {\it Astropart. Phys} {\bf 9}, 105 (1998).
\bibitem{wimp}       M.Ambrosio et al., {\it Phys. Rev.} {\bf D60}, 082002 (1999).
\bibitem{wimpnew}    T.Montaruli et al., HEP-EX/9905021 (1999).
\bibitem{susywimp}   A.Bottino et al., {\it Phys. Rev.} {\bf D62}, 056006 (2000).
\bibitem{summary}    J. Edsjo, {\it ``Review of indirect searches''} , this Conference.
%
\end{thebibliography}
\end{document}